\documentclass[onecollarge]{svjour2}
\usepackage{amsfonts}
\usepackage{amssymb}
\usepackage{amsmath}
\usepackage{graphicx}
\newcommand{\bm}{\mathbf} 
\journalname{Few-Body Systems}
\begin{document}

\title{Six-bodies calculations using the Hyperspherical Harmonics method
\thanks{Presented at the Sixth Workshop on the Critical Stability of Quantum
Few-Body Systems, Erice, Sicily, October 2011}}


\author{M. Gattobigio  \and A. Kievsky \and M. Viviani}

\authorrunning{Gattobigio et al.} 

\institute{ M. Gattobigio \at 
Universit\'e de Nice-Sophia Antipolis, Institut Non-Lin\'eaire de Nice, CNRS,
1361 route des Lucioles, 06560 Valbonne, France  \\
  \email{mario.gattobigio@inln.cnrs.fr}
            \and
A. Kievsky - M. Viviani \at
Istituto Nazionale di Fisica Nucleare,
Largo Pontecorvo 2, 56127 Pisa, Italy 
}

\date{Received: date / Accepted: date}

\maketitle

\begin{abstract}

In this work we show results for light nuclear systems and
small clusters of helium atoms using the hyperspherical harmonics basis. 
We use the basis without previous symmetrization or antisymmetrization
of the state. After the diagonalization of the Hamiltonian matrix, the
eigenvectors have well defined symmetry under particle permutation and the
identification of the physical states is possible.
We show results for systems composed up to six particles.
As an example of a fermionic system, we consider a nucleon system interacting
through the Volkov potential, used many times in the literature. 
For the case of bosons, we consider helium atoms interacting through a
potential model which does not present a strong repulsion at
short distances. We have used an attractive gaussian potential to reproduce the
values of the dimer binding energy, the atom-atom scattering length, and the
effective range obtained with one of the most widely used He-He interaction, the
LM2M2 potential.  In addition, we include a repulsive hypercentral
three-body force to reproduce the trimer binding energy. 

\end{abstract}

\section{Introduction}

The Harmonic Hyperspherical (HH) method is extensively used in the description
of few-body systems. For example the HH method has been applied to describe
bound states of $A=3,4$ nuclei (for a recent review see
Refs.~\cite{kievsky:1997_few-bodysyst,kievsky:2008_j.phys.g}).  In these applications the HH basis
elements, extended to spin and isospin degrees of freedom, have been combined
in order to construct antisymmetric basis functions; in fact, the HH functions,
as normally defined, do not have well defined properties under particle
permutation, but several schemes have been proposed to construct HH
functions with an arbitrary permutational symmetry, see 
Refs.~\cite{novoselsky:1994_phys.rev.a,%
novoselsky:1995_phys.rev.a,barnea:1999_phys.rev.a,timofeyuk:2008_phys.rev.c}. 

All of the proposed symmetrization schemes share an increasing computational
difficulty as the number of particles $A$ increases; to cope with this issue,
the authors proposed in Ref.~\cite{gattobigio:2009_phys.rev.a} to renounce to
the symmetrization step. If the Hamiltonian commutes with the group of
permutations of $A$ objects, $S_A$, the eigenvectors 
can be organized in accordance with the irreducible
representations of $S_A$; in fact, if there is no more degenerancy, the eigenvectors
have a well defined permutation symmetry. After the identification of the eigenvectors
belonging to the desired symmetry, the corresponding energies are variational
estimates.  The disadvantage of this method results in the large dimension of
the matrices to be diagonalized. However, at present, different techniques are
available to treat (at least partially) this problem. 

In order to show the main characteristics of this method,  we will discuss
results for bound states up to six particles interacting through a central
potential in two different systems: (i) a nucleon system interacting {\em via}
the Volkov potential, used many times in the
literature~\cite{barnea:1999_phys.rev.a,varga:1995_phys.rev.c,timofeyuk:2002_phys.rev.c,%
viviani:2005_phys.rev.c,gattobigio:2011_phys.rev.c,gattobigio:2011_j.phys.:conf.ser.},
and thus useful to test our approach, and (ii) a systems composed by helium
atoms interacting through a soft-core potential. The {\em ab-initio}-helium
potentials have a strong repulsion at small distances which makes calculations
quite difficult; few calculations exist on clusters of helium with these
potentials~\cite{lewerenz:1997_j.chem.phys.,blume:2000_j.chem.phys.,hiyama:2012_phys.rev.a}. On the
other hand, descriptions of few-atoms systems using soft-core potentials are
currently operated (see for example
Ref.~\cite{von_stecher:2009_phys.rev.a,kievsky:2011_few-bodysyst,gattobigio:2011_phys.rev.a}).

The paper is organized as follows. In Sect.~\ref{sec:method} a brief description of
the method is given. In Sect.\ref{sec:applications}, applications of the method
to a system of nucleons and helium atoms are shown. The conclusions are given
in the Sect.~\ref{sec:conclusions}.

\section{The unsymmetrized HH expansion}\label{sec:method}

In the present section we give a brief description of the HH basis showing
some properties of the basis that allow to use unsymmetrized basis
elements to describe a system of identical particles.

\subsection{The HH basis set}

Following Refs.\cite{gattobigio:2011_phys.rev.c,gattobigio:2011_phys.rev.a,gattobigio:2009_phys.rev.a},
we start with the definition of the Jacobi
coordinates for an equal mass $A$ body system,
with Cartesian coordinates $\mathbf r_1 \dots \mathbf r_A$

\begin{equation}
  \mathbf x_{N-j+1} = \sqrt{\frac{2 j}{j+1} } \,
                  (\mathbf r_{j+1} - \mathbf X_j)\,,
   \qquad
   j=1,\dots,N\,.
  \label{eq:jc2}
\end{equation}
with $\mathbf X_j = \sum_{i=1}^j \mathbf r_{j}/j$.
For a given set of Jacobi coordinates $\mathbf x_1, \dots, \mathbf x_N$, 
we can introduce the hyperspherical coordinates. A useful tool to represent 
hyperspherical coordinates  is the hyperspherical 
tree. This is a rooted-binary tree whose leaves represent the modules of Jacobi coordinates.
Once we introduce the hyperradius,
\begin{equation}
  \rho = \bigg(\sum_{i=1}^N x_i^2\bigg)^{1/2}
   = \bigg(2\sum_{i=1}^A (\mathbf r_i - \mathbf X)^2\bigg)^{1/2}
   = \bigg(\frac{2}{A}\sum_{j>i}^A (\mathbf r_j - \mathbf r_i)^2\bigg)^{1/2} \,,
  \label{}
\end{equation}
the modulus of the Jacobi coordinates live in a $(N-1)$-sphere of radius $\rho$ and
we can introduce $N-1$ hyperangles  to express the Jacobi coordinates as a function
of the hyperradius. The choice is not unique, and different choices are
represented by different hyperspherical trees~\cite{kildyushov:1972_sov.j.nucl.phys.,kildyushov:1973_sov.j.nucl.phys.}.
The relation between  a tree and the corresponding Jacobi coordinates is the
following; for each tree's node, labelled by $a$, we have an hyperangle
$\phi_a$. The rule to reconstruct the value of a Jacobi coordinate modululs reads: start
from the root node, and look for the path leading to the leaf corresponding to
the Jacobi coordinate; for each branch turning toward the left (right) we
multiply the hyperradius $\rho$ for cosine (sine) of the hyperangle attached to
the branching point. As an example, in Eq.~(\ref{eq:A5_nonStandard}) we have a
tree choice for $A=5$ with the corresponding relations between Jacobi and hyperspherical
coordinates
\begin{equation}
  \begin{aligned}
  x_1 &= \rho \sin\phi_4\sin\phi_2 \\
  x_2 &= \rho \sin\phi_4\cos\phi_2 \\
  x_3 &= \rho \cos\phi_4\sin\phi_3 \\
  x_4 &= \rho \cos\phi_4\cos\phi_3 \,, \\
  \end{aligned}\quad\quad\quad
   \begin{minipage}{0.25\linewidth}
  \includegraphics[width=\linewidth]{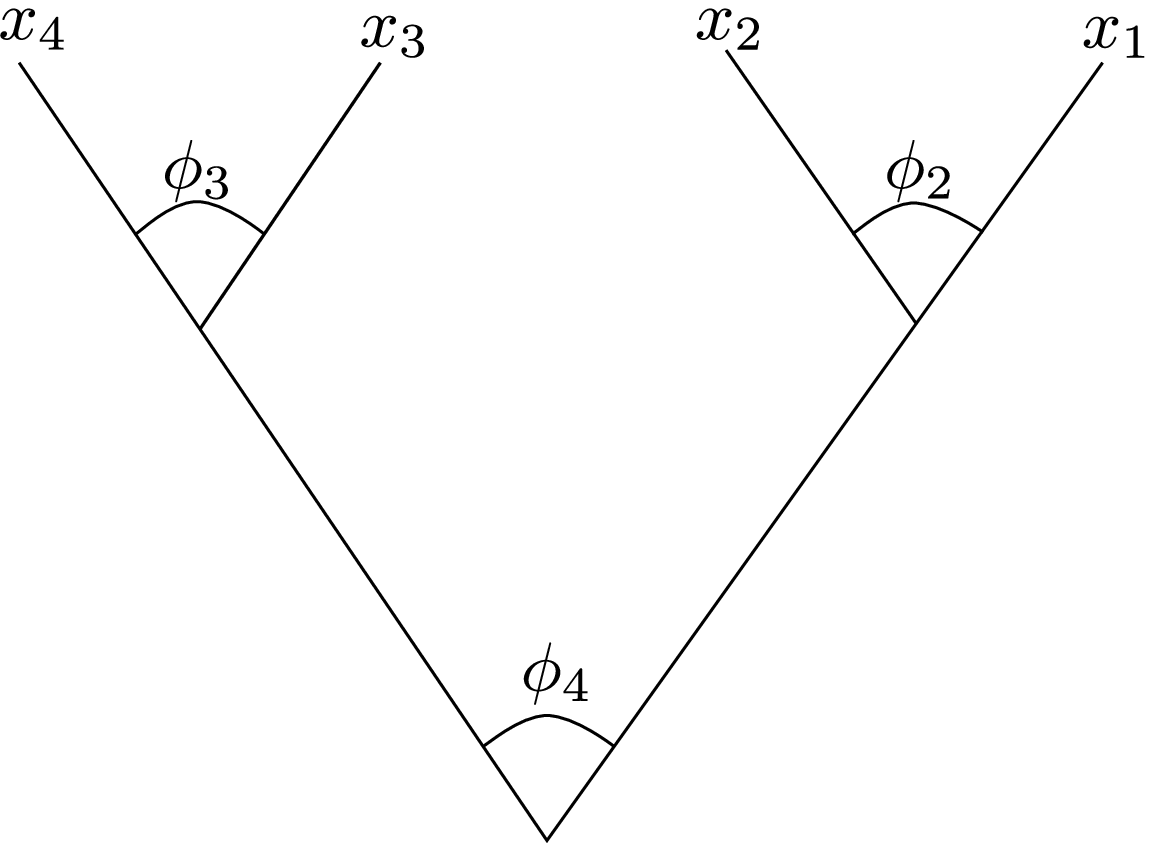}
   \end{minipage} \,.
  \label{eq:A5_nonStandard}
\end{equation}
Different trees have different topologies; given a node $a$,
the left (right) branch connects the node to a sub-binary tree made up
of $N_a^{l(r)}$ nodes and $L_a^{l(r)}$ leaves. We can use this information to
construct useful topological numbers as
\begin{equation}
  C_a = N_a^l + \frac{1}{2} L_a^l + \frac{1}{2}\,,
  \label{eq:CtopologicalNumber}
\end{equation}
and
\begin{equation}
  S_a = N_a^r + \frac{1}{2} L_a^r + \frac{1}{2}\,.
  \label{eq:StopologicalNumber}
\end{equation}

\begin{figure}[t]
  \begin{center}
 \includegraphics[width=0.35\linewidth]{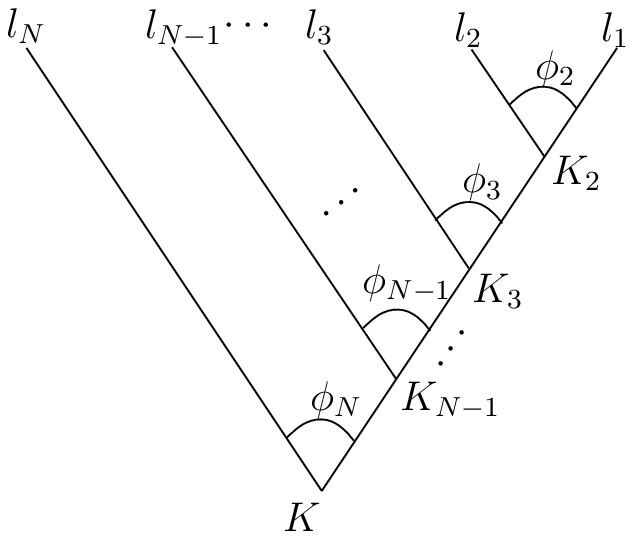}%
 \hspace{2cm}%
 \includegraphics[width=0.35\linewidth]{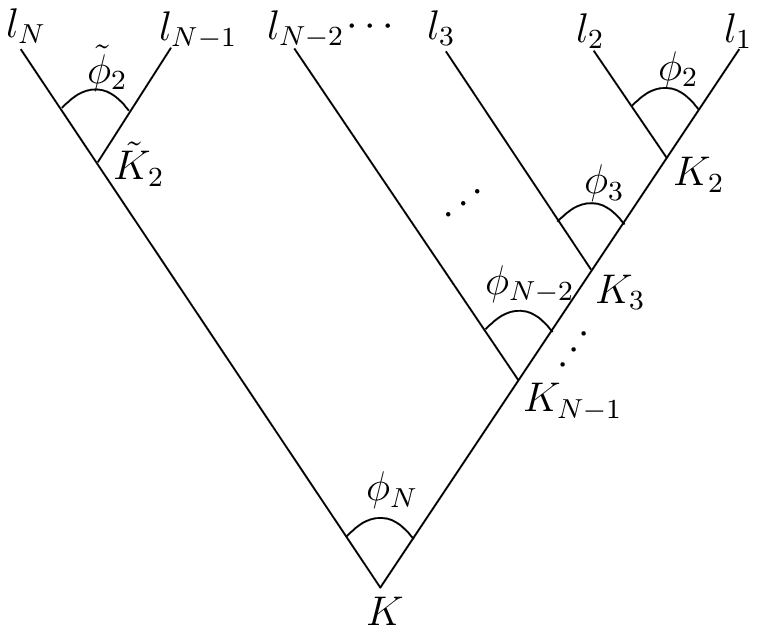}
  \end{center}
  \caption{In the left panel we have drawn the standard hyperspherical tree;
    this is the one used in the standard definition of the basis, and the one
    used to calculate the two-body potential between particles at $\mathbf r_1$
    and $\mathbf r_2$.  In the right panel we have drawn the non-standard tree,
  used to calculate the three-body force between particles at $\mathbf r_1$,
$\mathbf r_2$, and $\mathbf r_3$.}
 \label{fig:tree}
\end{figure}

The set of the hyperangles together with the direction of the Jacobi coordinates
$\hat{\mathbf x}_i =(\varphi_i,\theta_i)$ form the hyperangular coordinates 
\begin{equation}
  \Omega_N = (\hat {\bm x}_1, \dots, \hat {\bm x}_N, \phi_2, \dots, \phi_N) \,.
  \label{}
\end{equation}
in terms of which the HH functions ${\mathcal
Y}_{[K]}(\Omega_N)$ are defined.
The subscript
$[K]$ stands for the set of $(3N-1)$-quantum numbers $l_1,\dots,l_N,m_1, \dots,m_N,
K_2, \dots, K_N$, with $K_N=K$ the grand-angular momentum.
They can be expressed
in terms of the usual harmonic functions $Y_{lm}(\hat {\bm x})$ and of the
Jacobi polynomials $P_n^{a,b}(z)$
\begin{equation}
  {\mathcal Y}_{[K]}^{LM}(\Omega_N) = 
    \left[\prod_{j=1}^N Y_{l_jm_j}(\hat {\bm x}_j) \right]_{LM} 
    \left[ \prod_{a\in\text{nodes}}
    {\mathcal P}_{K_a}^{\alpha_{K_a^l},\alpha_{K_{a}^{r\phantom{l}}}}(\phi_a)\right] \,,
  \label{eq:hh}
\end{equation}
with
\begin{equation}
    {\mathcal P}_{K_a}^{\alpha_{K_a^l},\alpha_{K_{a}^{r\phantom{l}}}}(\phi_a)
    =
{\mathcal
N}_{n_a}^{\alpha_{K_{a}^{r\phantom{l}}},\alpha_{K_a^l}} 
(\cos\phi_a)^{K_a^l} (\sin\phi_a)^{K_a^r} 
P^{\alpha_{K_{a}^{r\phantom{l}}},\alpha_{K_a^l}}_{n_a}(\cos2\phi_a) \,,
\end{equation}
where we have defined 
\begin{equation}
  \alpha_{K_a^{l(r)}} = K_a^{l(r)} + N_a^{l(r)} + \frac{1}{2} L_a^{l(r)}\, .
  \label{}
\end{equation}
The normalization factor reads
\begin{equation}
  {\cal N}_{n}^{\alpha\beta} =
  \sqrt{\frac{2(2n+\alpha+\beta+1) n!\,
  \Gamma(n+\alpha+\beta+1)}{\Gamma(n+\alpha+1)\Gamma(n+\beta+1)}}\,.
  \label{eq:norma}
\end{equation}
With the above definitions, the HH functions have well defined total
orbital angular momentum $L$ and $z$-projection $M$. 
The standard choice of hyperspherical coordinates, and of the 
corresponding HH, is represented in the left panel of 
Fig.~\ref{fig:tree}; this is the one we use as our basis set.

\subsection{Rotation matrices between HH basis elements of different Jacobi coordinates}

Here we are interested in a particular set of coefficients relating the 
reference HH basis to a basis in which the ordering of two adjacent
particles have been transposed. 
In the transposition between particles $j,j+1$, only the Jacobi vectors
$\mathbf x_i$ and $\mathbf x_{i+1}$, with $i=N-j+1$, are different. 
We label them $\mathbf x'_i$ and $\mathbf x'_{i+1}$, and explicitly they are
\begin{equation}
  \begin{aligned}
 \mathbf x'_{i}  &= - \frac{1}{j} \,\mathbf x_i + 
                      \frac{\sqrt{(j+1)^2-2(j+1)}}{j}  \,\mathbf x_{i+1} \\
 \mathbf x'_{i+1}&=   \frac{\sqrt{(j+1)^2-2(j+1)}}{j} \,\mathbf x_i 
                    + \frac{1}{j} \,\mathbf x_{i+1}   \,,
  \end{aligned}
  \label{eq:jc3}
\end{equation}
with $i=1,\ldots,N-1$. The corresponding moduli verify
${x'}^2_i+{x'}^2_{i+1}=x^2_i+x^2_{i+1}$. 
Let us call ${\mathcal Y}^{LM}_{[K]}(\Omega^i_N)$ the HH basis element
constructed in terms of a set of Jacobi coordinates in which the $i$-th
and $i+1$-th Jacobi vectors are given from Eq.(~\ref{eq:jc3}) with all the
other vectors equal to the original ones (transposed basis). 
The coefficients
\begin{equation}
 {\mathcal A}^{i,LM}_{[K][K']}=\int d\Omega_N[{\mathcal Y}^{LM}_{[K]} (\Omega_N)]^*
 {\mathcal Y}^{LM}_{[K']}(\Omega^i_N)\,,
\label{eq:ca1}
\end{equation}
are the matrix elements of a matrix ${\mathcal A}^{LM}_i$
 that allows to express the transposed HH basis 
elements in terms of the reference basis. 
The total angular momentum as well as the grand angular quantum number $K$
are conserved in the above integral ($K=K'$). 
The coefficients ${\mathcal A}^{i,LM}_{[K][K']}$ form a very sparse matrix and
they can be calculated analytically using angular and 
${\cal T}$-coupling coefficients (Kil'dyushov coefficients)
and the Raynal-Revai matrix elements~\cite{gattobigio:2011_phys.rev.c,gattobigio:2011_few-bodysyst.,%
gattobigio:2011_phys.rev.a} .

We are now interested in obtaining the rotation coefficients between the
reference HH basis and a basis in which the last
Jacobi vector is defined as $\mathbf x'_N=\mathbf r_j-\mathbf r_i$,
without loosing generality we consider $j>i$. 
A generic rotation coefficient of this kind can be
constructed as successive products of the ${\mathcal A}^{k,LM}_{[K][K']}$
coefficients. 
Defining ${\mathcal Y}^{LM}_{[K]}(\Omega^{ij}_N)$ the HH basis element
constructed in terms of a set of Jacobi coordinates in which the 
$N$-th Jacobi vector is defined $\mathbf x'_N=\mathbf r_j-\mathbf r_i$,
the rotation coefficient relating this basis to the reference basis
can be given in the following form
\begin{equation}
 {\mathcal B}^{ij,LM}_{[K][K']}=\int d\Omega[{\mathcal Y}^{LM}_{[K]} (\Omega_N)]^*
 {\mathcal Y}^{LM}_{[K]}(\Omega^{ij}_N) =
\left[{\mathcal A}^{LM}_{i_1}\cdots{\mathcal A}^{LM}_{i_n}\right]_{[K][K']} \,.
\label{eq:ca2}
\end{equation}
The particular values of the indices $i_1,\ldots,i_n$, labelling
the matrices ${\mathcal A}^{LM}_{i_1},\ldots,{\mathcal A}^{LM}_{i_n}$, 
depend on the pair $(i,j)$.
The number of factors cannot be greater than $2(j-2)$ and it increases,
at maximum, by two units from $j$ to $j+1$. 
The matrix
\begin{equation}
{\mathcal B}_{ij}^{LM}={\mathcal A}^{LM}_{i_1}\cdots{\mathcal A}^{LM}_{i_n}\,,
\label{eq:matrixb}
\end{equation}
is written as a product of the sparse matrices ${\mathcal A}^{LM}_{i}$'s, 
a property which 
is particularly well suited for a numerical implementation of the
potential energy matrix.

\subsection{The two-body and three-body potential energy matrices}\label{sec:pot}

We consider the potential energy of an $A$-body system constructed in terms 
of two-body interactions
\begin{equation}
   V=\sum_{i<j} V(i,j)  \;\;\; .
\end{equation}
In the case of a central two-body interaction, its matrix 
elements in terms of the HH basis are
\begin{equation}
   V_{[K][K']}(\rho)=\sum_{i<j} 
\langle{\cal Y}^{LM}_{[K]}(\Omega_N)|V(i,j)|{\cal
Y}^{LM}_{[K']}(\Omega_N)\rangle \, .
\end{equation}
In each element $\langle{\cal Y}^{LM}_{[K]}|V(i,j)|{\cal Y}^{LM}_{[K']}\rangle$ the integral
is understood on all the hyperangular variables and depends parametrically on 
$\rho$. Explicitly, for the pair $(1,2)$, the matrix elements of the matrix 
$V_{12}(\rho)$ are
\begin{equation}
\begin{aligned}
& V^{(1,2)}_{[K][K']}(\rho)=
\langle{\cal Y}^{LM}_{[K]}(\Omega_N)|V(1,2)|{\cal Y}^{LM}_{[K']}(\Omega_N)\rangle= \cr
&\delta_{l_1,l^\prime_1}\cdots\delta_{l_N,l^\prime_N}
\delta_{L_2,L^\prime_2}\cdots\delta_{L_N,L^\prime_N}
\delta_{K_2,K^\prime_2}\cdots\delta_{K_N,K^\prime_N} \cr
&\times \int d\phi_N(\cos\phi_N\sin\phi_N)^2
\;{}^{(N)}{\cal P}^{l_N,K_{N-1}}_{K_N}(\phi_N)
V(\rho\cos\phi_N)\;{}^{(N)}{\cal P}^{l_N,K_{N-1}}_{K'_N}(\phi_N)\,.
\end{aligned}
\label{eq:v12}
\end{equation}
Using the rotation coefficients, a general term of the potential $V(i,j)$ results
\begin{equation}
 V^{(i,j)}_{[K][K']}(\rho)=
\sum_{[K''][K''']}{\cal B}^{ij,LM}_{[K''][K]}{\cal B}^{ij,LM}_{[K'''][K']}
\langle{\cal Y}^{LM}_{[K'']}(\Omega^{ij}_N)|V(i,j)|{\cal
Y}^{LM}_{[K''']}(\Omega^{ij}_N)\rangle \,.
\label{eq:vij}
\end{equation}
or, in matrix notation,
\begin{equation}
  V_{ij}(\rho)= [{\cal B}^{LM}_{ij}]^{t} \,V_{12}(\rho)\,{\cal B}^{LM}_{ij}\,.
\label{eq:mij}
\end{equation}
The complete potential matrix energy results
\begin{equation}
\sum_{i<j} V_{ij}(\rho)=\sum_{i<j} 
[{\cal B}^{LM}_{ij}]^t\, V_{12}(\rho)\,{\cal B}^{LM}_{ij} \,.
\label{eq:vpot}
\end{equation}

Each term of the sum in Eq.(\ref{eq:vpot}) results in a product of sparse
matrices, a property which allows an efficient implementation of matrix-vector
product, key ingredient in the solution of the Schr\"odinger equation using
iterative methods.

The three-body force used in the present work depends on the hyperradius
$\rho_{ijk}$ of a triplet of particles $\mathbf r_i,\mathbf r_j,\mathbf r_k$.
For an $A$-body systems, there are $\binom{A}{3}$ three-body terms
\begin{equation}
  V^{(3)} = \sum_{i<j<k} W(\rho_{ijk})\,,
  \label{}
\end{equation}
and one of them is the force between the triplet $\mathbf r_1,\mathbf
r_2,\mathbf r_3$ for which we have $\rho^2_{123} = x^2_N+x^2_{N-1}$. This term
can be easily calculated on a hyperspherical-basis set relative to an
non-standard hyperspherical tree with the branches attached to leaves $x_N$ and
$x_{N-1}$ going to the same node, as shown in the right panel of
Fig.~\ref{fig:tree}.  The transition between this tree and the standard tree is
simply given by the ${\cal T}$-coefficients
\begin{equation}
   \begin{minipage}{3.5cm}
  \includegraphics[width=\linewidth]{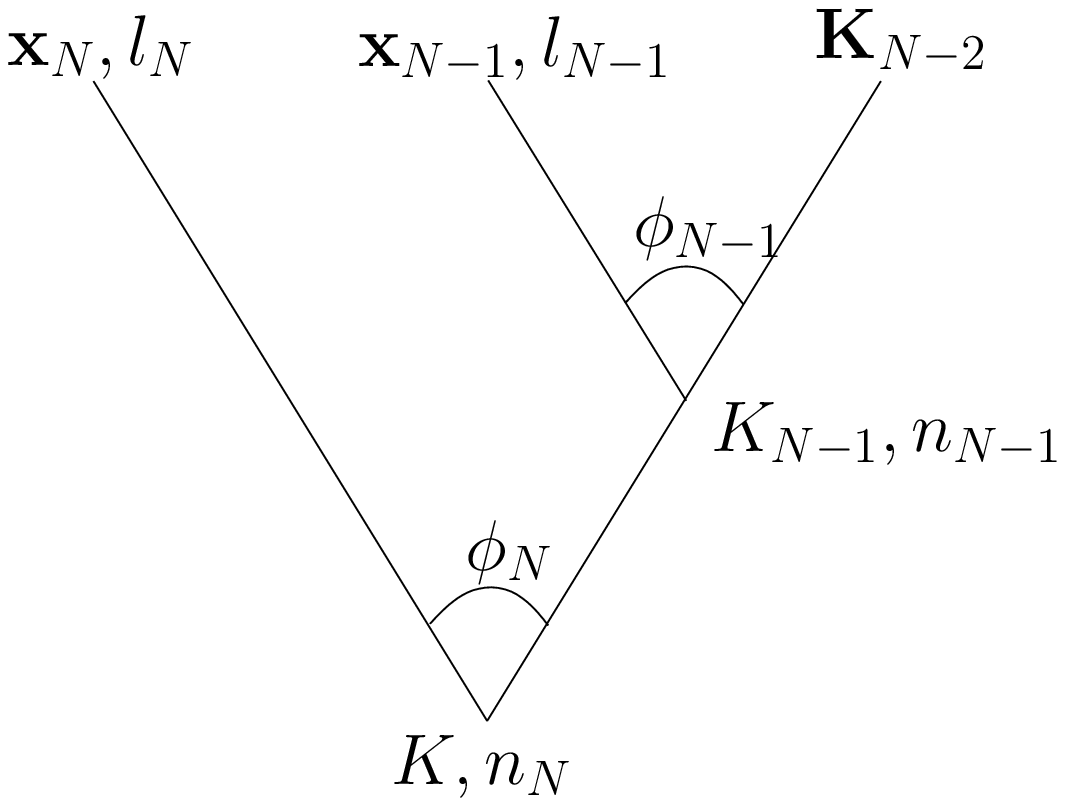}
   \end{minipage} 
   =
   \sum_{\tilde n_{N-1}}
   {\cal T}^{\alpha_{K_{N-2}}\alpha_{l_{N-1}}\alpha_{l_N}}_{n_{N-1} \;\tilde
   n_{N-1}\; K}
   \begin{minipage}{3.5cm}
  \includegraphics[width=\linewidth]{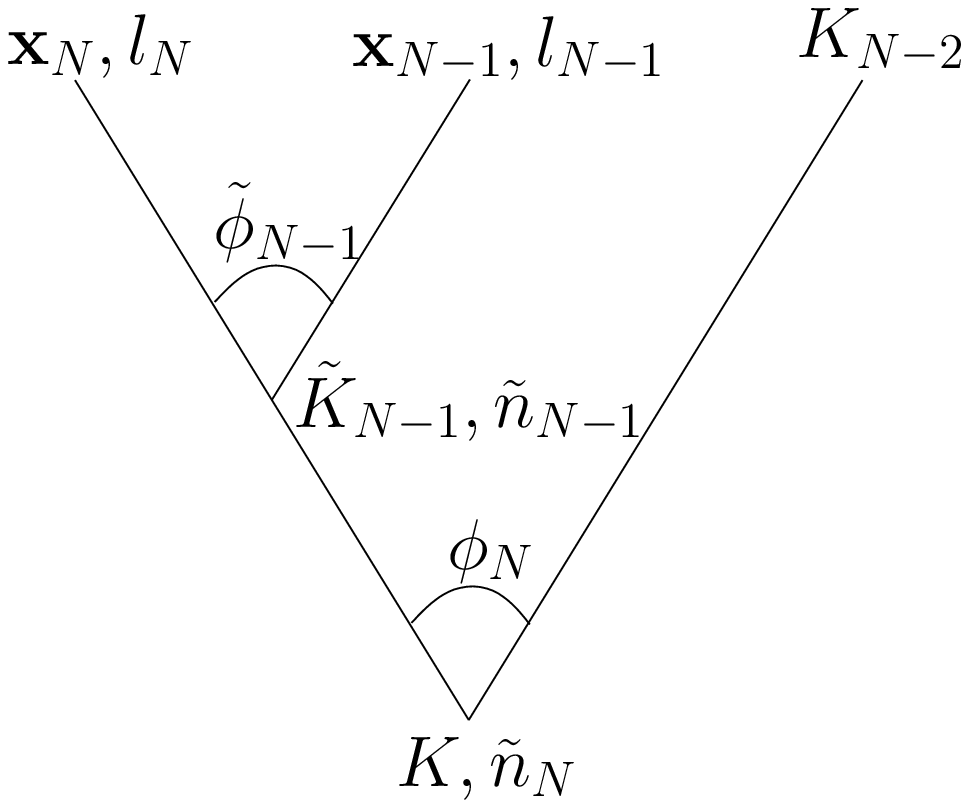}~\,,
   \end{minipage} 
\end{equation}
or
\begin{equation}
  {\cal Y}^{LM}_{[K]}(\Omega_N)
   =
   \sum_{\tilde n_{N-1}} 
   {\cal T}^{\alpha_{K_{N-2}}\alpha_{l_{N-1}}\alpha_{l_N}}_{n_{N-1} \;\tilde
   n_{N-1}\; K}
   {\cal Y}^{LM}_{[\tilde K]}(\tilde\Omega_N) \,,
  \label{}
\end{equation}
where all the variable with the tilde refer to the non-standard tree.
In fact, with this choice we simply have
\begin{equation}
  \rho_{123} = \rho\cos\phi_N\,,
  \label{}
\end{equation}
and the fixed-rho matrix elements of the matrix $W_{123}(\rho)$ are
\begin{equation}
  \begin{aligned}
&    \langle {\cal Y}^{LM}_{[\tilde K']}(\tilde\Omega_N) | W(\rho_{123}) | {\cal Y}^{LM}_{[\tilde K]}(\tilde\Omega_N) \rangle
  =  \cr
&  \delta_{l'_1,l_1} \cdots \delta_{l'_N,l_N}
  \delta_{L'_2,L_2} \cdots \delta_{L',L} \delta_{M',M}
  \delta_{\tilde K'_2,\tilde K_2} \cdots \delta_{\tilde K'_{N-1},\tilde K_{N-1}}\times  \cr
&  \int (\cos\phi_N)^{C_K}(\sin\phi_N)^{S_K} d\phi_N\;
  {\mathcal P}_{K'}^{\alpha_{\tilde K_{N-1}},\alpha_{K_{N-2}}}(\phi_N)  
  {\mathcal P}_{K}^{\alpha_{\tilde K_{N-1}},\alpha_{K_{N-2}}}(\phi_N) 
  W(\rho\cos\phi_N) \,,
  \label{}
  \end{aligned}
\end{equation}
where $C_K$ and $S_K$ are the topological quantum numbers 
relative to the grand-angular-$K$ root node.  In practice the matrix is extremely
sparse, and it is diagonal on all quantum numbers but the grand-angular
momentum. 

The three-body force matrix in the standard basis is obtained by means of 
the ${\cal T}$-coefficients
\begin{equation}
  \begin{aligned}
&    \langle {\cal Y}^{LM}_{[K']}(\Omega_N) | W(\rho_{123}) | {\cal Y}^{LM}_{[K]}(\Omega_N) \rangle
  = \cr
  & \sum_{\tilde n_{N-1}} 
   {\cal T}^{\alpha_{K_{N-2}}\alpha_{l_{N-1}}\alpha_{l_N}}_{n'_{N-1} \;\tilde
   n_{N-1}\; K'}
   {\cal T}^{\alpha_{K_{N-2}}\alpha_{l_{N-1}}\alpha_{l_N}}_{n_{N-1} \;\tilde
   n_{N-1}\; K}
    \langle {\cal Y}^{LM}_{[\tilde K']}(\tilde\Omega_N) | W(\rho_{123}) | {\cal
    Y}^{LM}_{[\tilde K]}(\tilde\Omega_N) \rangle \,,
  \end{aligned}
\end{equation}
which for all practical purposes reduces to a product of sparse matrices.

In order to calculate the other terms of the three-body force, we use the 
matrices ${\cal A}_p^{LM}$, defined in Eq.~(\ref{eq:ca1}), that transpose
particles; with a suitable product of these sparse matrices
\begin{equation}
  {\cal D}_{ijk}^{LM} =  {\cal A}_{p_1}^{LM} \cdots {\cal A}_{p_m}^{LM}\,,
  \label{}
\end{equation}
we can permute the
particles in such a way that $\mathbf x_N = \mathbf r_i-\mathbf r_j$, 
and $\mathbf x_{N-1} = 2/\sqrt{3}(\mathbf r_k - (\mathbf r_i+\mathbf r_j)/2)$,
and $\rho^2_{ijk} = x_{N-1}^2 +  x_N^2$, and the total three-body force reads
\begin{equation}
  V^{(3)} = \sum_{i<j<k} [{\cal D}_{ijk}^{LM}]^t\, W_{123}(\rho) \,
  {\cal D}_{ijk}^{LM}  \,.
  \label{}
\end{equation}

\section{Applications of the HH expansion up to six particles}\label{sec:applications}

In this section we present results for $A=3-6$ systems obtained by
a direct diagonalization of the Hamiltonian of the system. The
corresponding Hamiltonian matrix is obtained using
the following orthonormal basis
\begin{equation}
  \langle\rho\,\Omega\,|\,m\,[K]\rangle =
  \bigg(\beta^{(\alpha+1)/2}\sqrt{\frac{m!}{(\alpha+m)!}}\,
  L^{(\alpha)}_m(\beta\rho)
  \,{\text e}^{-\beta\rho/2}\bigg)
  {\cal Y}^{LM}_{[K]}(\Omega_N)  \,,
  \label{mhbasis}
\end{equation}
where $L^{(\alpha)}_m(\beta\rho)$ is a Laguerre polynomial with
$\alpha=3N-1$ and $\beta$ a variational non-linear parameter.
The matrix elements of the Hamiltonian are obtained after
integrations in the $\rho,\Omega$ spaces. They depend on
the indices $m,m'$ and $[K],[K']$ as follows
\begin{equation}
\begin{aligned}
  \langle m'\,[K']|H|\,m\,[K] \rangle = -\frac{\hbar^2\beta^2}{m}
 ( T^{(1)}_{m'm}-K(K+3N-2) T^{(2)}_{m'm}) \delta_{[K'][K]} \cr
 + \sum_{i<j} \left[
\sum_{[K''][K''']}{\cal B}^{ij,LM}_{[K][K'']}{\cal B}^{ij,LM}_{[K'''][K']}
 V^{m,m'}_{[K''][K''']}\right] \cr
 + \sum_{i<j<k} \left[
\sum_{[K''][K''']}{\cal D}^{ijk,LM}_{[K][K'']}{\cal D}^{ijk,LM}_{[K'''][K']}
 W^{m,m'}_{[K''][K''']}\right] \cr
 \,.
\end{aligned}
\label{eq:hmm}
\end{equation}
The matrices $T^{(1)}$ and $T^{(2)}$ have an analytical form
and are given in Ref.~\cite{gattobigio:2011_phys.rev.c}.  The matrix elements
$V^{m,m'}_{[K][K']}$ are obtained after integrating the matrix $V_{12}(\rho)$
in $\rho$-space whereas the matrix elements $W^{m,m'}_{[K][K']}$  are
obtained after integration of the matrix $W_{123}(\rho)$ 
(we will call the corresponding matrices $V_{12}$ and $W_{123}$, respectively).
Introducing the diagonal matrix $D$ such that
$\langle [K']\,|\,D\, | [K]\rangle = \delta_{[K],[K']} K(K+3N-2)$, and the identity
matrix $I$ in $K$-space, we can rewrite the Hamiltonian schematically as
\begin{equation}
  H = -\frac{\hbar^2\beta^2}{m} ({}^{(1)}T \otimes I  +  {}^{(2)}T\otimes D )
  + \sum_{ij} [{\cal B}^{LM}_{ij}]^t\, V_{12}\,{\cal B}^{LM}_{ij} 
  + \sum_{ijk} [{\cal D}^{LM}_{ijk}]^t\, W_{123}\,{\cal D}^{LM}_{kij} \,,
  \label{eq:schemH}
\end{equation}
in which the tensor product character of the kinetic energy is explicitly
given.

\subsection{nuclear system}\label{sec:appli_nucleons}
As a first example we consider e nuclear system interaction through  a
simple two-body potential, the Volkov potential
\begin{equation}
 V(r)=V_R \,{\rm e}^{-r^2/R^2_1} + V_A\, {\rm e}^{-r^2/R^2_2} \,,
\end{equation}
with $V_R=144.86$ MeV, $R_1=0.82$ fm, $V_A=-83.34$ MeV, and $R_2=1.6$ fm.
The nucleons are considered to have the same mass chosen to be equal to the
reference mass $m$ and corresponding to
$\hbar^2/m = 41.47~\text{MeV\,fm}^{2}$.
With this parametrization of the potential, the
two-nucleon system has a binding energy $E_{2N}=0.54592\;$MeV and a
scattering length $a_{2N}=10.082\;$fm.
This potential has been used several times in the literature making its
use very useful to compare different methods
\cite{barnea:1999_phys.rev.a,varga:1995_phys.rev.c,timofeyuk:2002_phys.rev.c,viviani:2005_phys.rev.c}. The use of central
potentials in general produces too much binding, in particular the
$A=5$ system results bounded. Conversely, the use of the
$s$-wave version of the potential produces a spectrum much closer
to the experimental situation. This is a direct consequence of the
weakness of the nuclear interaction in $p$-waves. Accordingly,
we analyze this version of the potential, the $s$-wave projected potential.
The results are obtained
after a direct diagonalization of the Hamiltonian matrix of
Eq.(\ref{eq:hmm}) including $m_{max}+1$ Laguerre polynomials with a fix
value of $\beta$, and all
HH states corresponding to maximum value of the grand angular momentum
$K_{max}$. The scale parameter $\beta$ can be used as a non-linear
parameter to study the convergence in the index $m=0,1,\ldots,m_{max}$, with
$m_{max}$ the maximum value considered.
We found that $20$ Laguerre polynomials (with proper
values of $\beta$) were sufficient for an accuracy of $0.1$\%
in the calculated eigenvalues.

\begin{figure}[t]
  \begin{center}
    \includegraphics[width=0.5\linewidth]{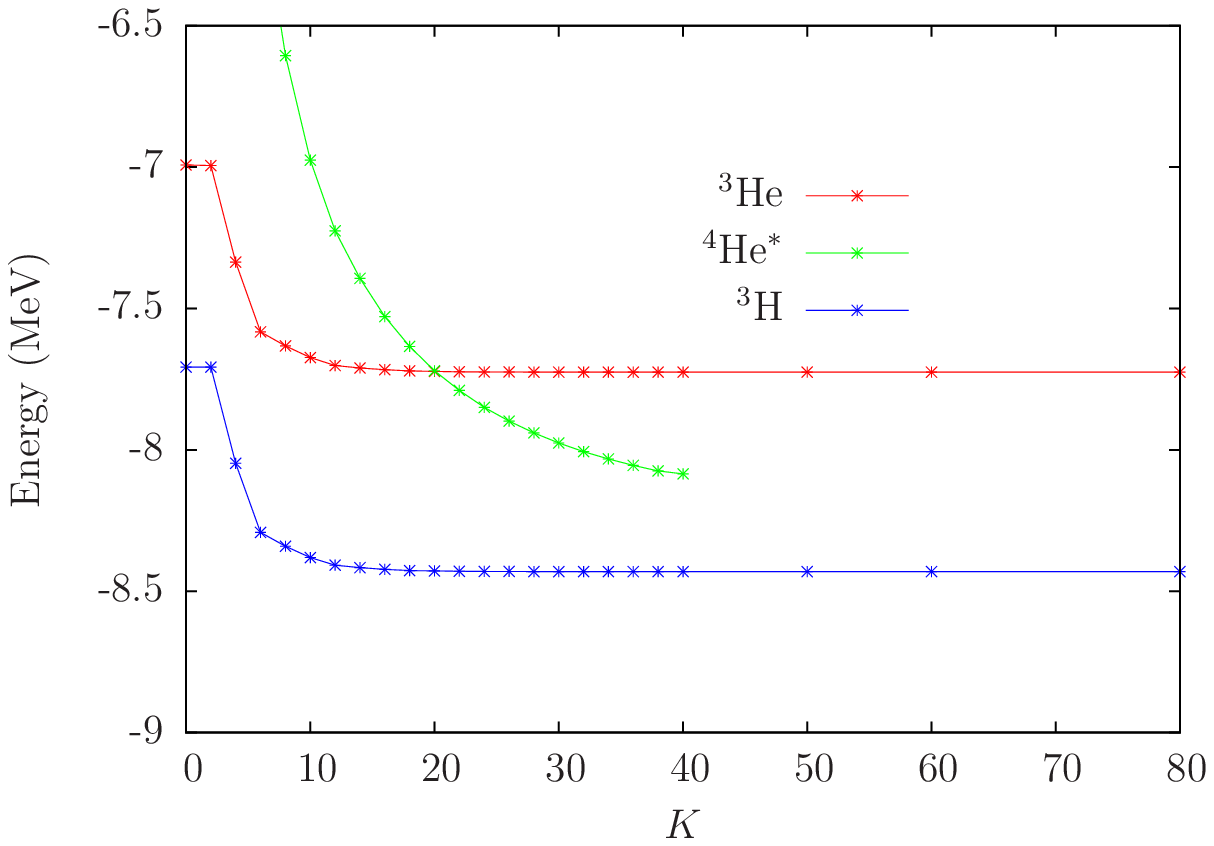}%
    \includegraphics[width=0.5\linewidth]{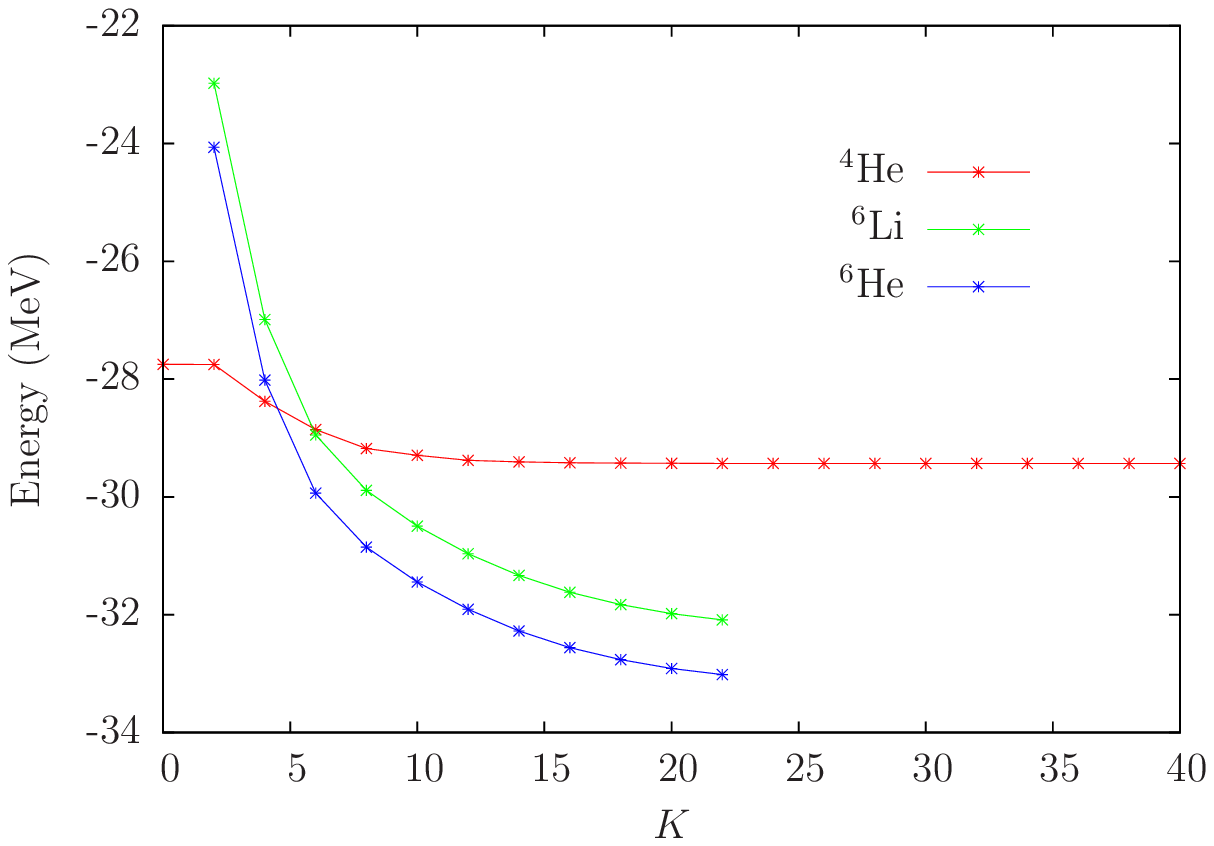}
  \end{center}
\caption{In the left panel we have the convergence of the $^3$H and $^3$He
  binding energies as a function of $K$. The excited state of the alpha
  particle $^4$He$^*$ is also shown.  In the right panel we have the
  convergence of the $^4$He, $^6$He and $^6$Li binding energies as a function
of $K$.}
\label{fig:convergences}
\end{figure}

The results of the present analysis are given in Fig.~\ref{fig:convergences}
where the convergence of the $A=3-6$ binding energies are given as a function
of $K$.  In the left panel of Fig.~\ref{fig:convergences} the convergence for
the excited state $^4$He$^*$ of the $\alpha$ particle is also shown.  For
$A=3,4$ a very extended HH expansion has been  used with the maximum value of
$K=80$ and $K=40$ respectively.  For $A=3$, the obtained results are 8.431 MeV
and 7.725 MeV for $^3$H and $^3$He respectively.  For $A=4$, the ground state
binding energy converges at the level of $1-2$~keV for $K_{max}=40$.  The
convergence of the excited state $^4$He$^*$ has been estimated at the level of
$50$~keV.  Though the convergence was not completely achieved, the description
is close to the experimental observation of a $0^+$ resonance between the two
thresholds and centered 395 keV above the $p$-$^3$H threshold.  Besides its
simplicity, the $s$-wave potential describes the $A=3,4$ system in reasonable
agreement with the experiment.

For the $A=6$ system a maximum value of $K=22$ has been used which greatly
improve previous attemps in using the HH basis in $A=6$ systems
\cite{novoselsky:1995_phys.rev.a,timofeyuk:2008_phys.rev.c}.  The obtained
results are 33.016 MeV and 32.087 MeV for $^6$He and $^6$Li respectively.  It
should be noticed that these states belong to the mixed symmetries
$[\bf{4}\,{\bf 2}]$ (without the Coulomb interaction).  When the Coulomb
interaction between two nucleons is included the state belong to the symmetry
$[\bf{2}]\otimes[{\bf 2}^2]$ and when it is included between three nucleons the
state belongs to the symmetry $[\bf{2}\,\bf{1}]\otimes[{\bf 2}\,\bf{1}]$.
These states are embedded in a very dense spectrum. In order to follow these
state in the projected Lanczos method a projection-purification procedure is
performed.

\subsection{atomic system}\label{sec:appli_atoms}

As an example of an atomic systems we describe a system of $^4$He atoms up to six atoms.
The $^4$He-$^4$He interaction presents a
strong repulsion at short distances, below 5 a.u. This characteristic makes
it difficult a detailed description of the system with more than four atoms.
Accordingly, we study small clusters of helium
interacting through a soft-core two- and three-body potentials. 
Following Refs.~\cite{nielsen:1998_j.phys.b,kievsky:2011_few-bodysyst,gattobigio:2011_phys.rev.a} 
we use the gaussian two-body potential
\begin{equation}
V(r)=V_0 \,\, {\rm e}^{-r^2/R^2}\,,
\label{twobp}
\end{equation}
with $V_0=-1.227$ K and $R=10.03$~a.u.. In the following we use
$\hbar^2/m=43.281307~\text{(a.u.)}^2\text{K}$. This parametrization of the
two-body potential approximately reproduces the dimer binding energy $E_{2}$,
the atom-atom scattering length $a_0$ and the effective range $r_0$ given by
the LM2M2 potential.  Specifically, the results for the gaussian potential are
$E_{2}=-1.296$ mK, $a_0=189.95$ a.u. and $r_0=13.85$ a.u., to be compared to
the corresponding LM2M2 values $E_{2}=-1.302$ mK, $a_0=189.05$ a.u. and
$r_0=13.84$ a.u..  As shown in Ref.~\cite{kievsky:2011_few-bodysyst}, the use of the
gaussian potential in the three-atom system produces a ground state binding
energy $E^{(0)}_{3}=150.4$ mK, which is appreciable bigger than the LM2M2 helium
trimer ground state binding energy of $126.4$ mK.  
In order to have a closer description to the $A=3$ system obtained with the
LM2M2 potential, we introduce the following three-body interaction
\begin{equation}
W(\rho_{ijk})=W_0 \,\, {\rm e}^{-2\rho^2_{ijk}/\rho^2_0}\,,
  \label{eq:hyptbf}
\end{equation}
where $\rho^2_{ijk}=\frac{2}{3}(r^2_{ij}+r^2_{jk}+r^2_{ki})$ is the three-body
hyperradius in terms of the distances of the three interacting particles.
Moreover, the strength $W_0$ is fixed to reproduce the LM2M2 helium trimer binding
energy of $126.4$ mK.  In Ref.~\cite{gattobigio:2011_phys.rev.a} a detailed analysis of this
force has been performed by varying the range $\rho_0$ between 4 and 16 a.u.. 
Here we present results for small clusters, up to $A=6$, formed by
atoms of $^4$He using the soft two-body force plus the hyperradial
three-body force with parameters $W_0,\rho_0\equiv 0.422 {\rm K},14.0$ a.u..  

\begin{figure}[h]
  \includegraphics[width=0.5\linewidth]{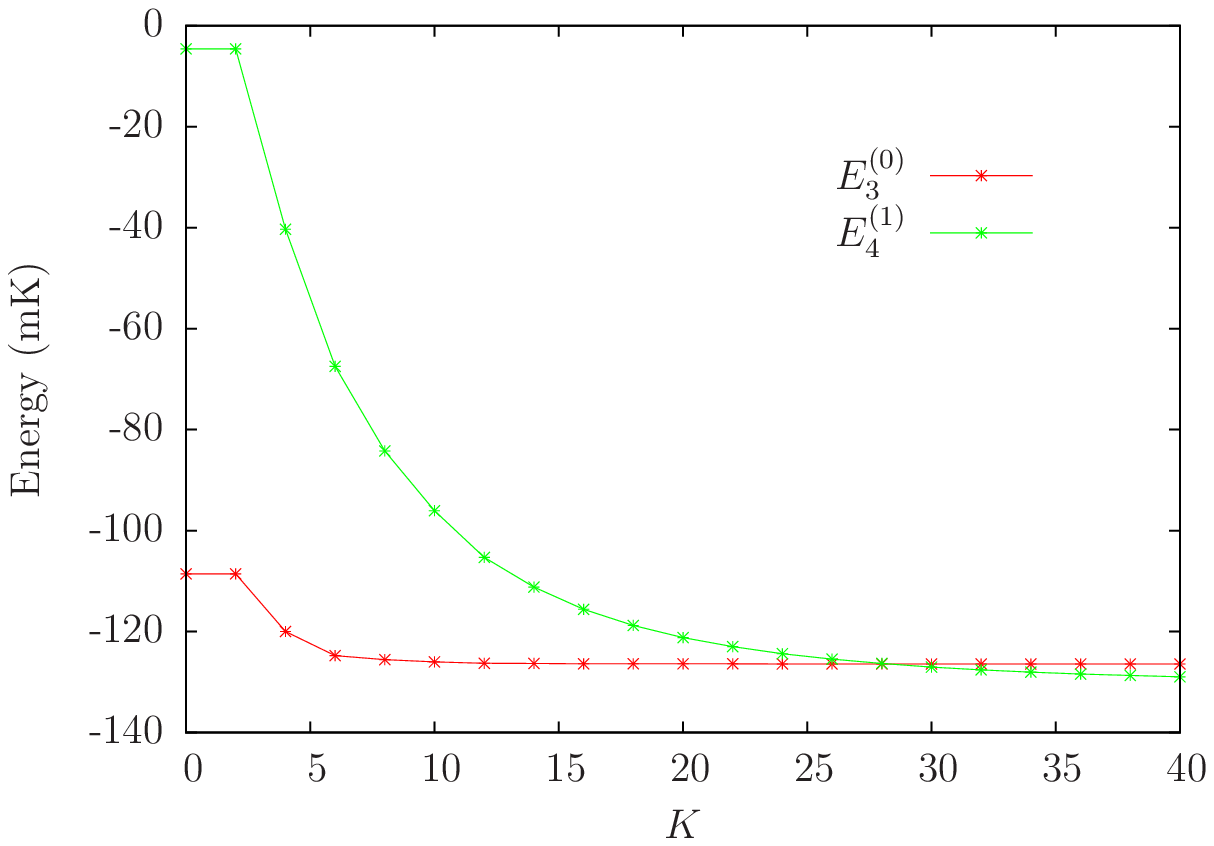}
  \includegraphics[width=0.5\linewidth]{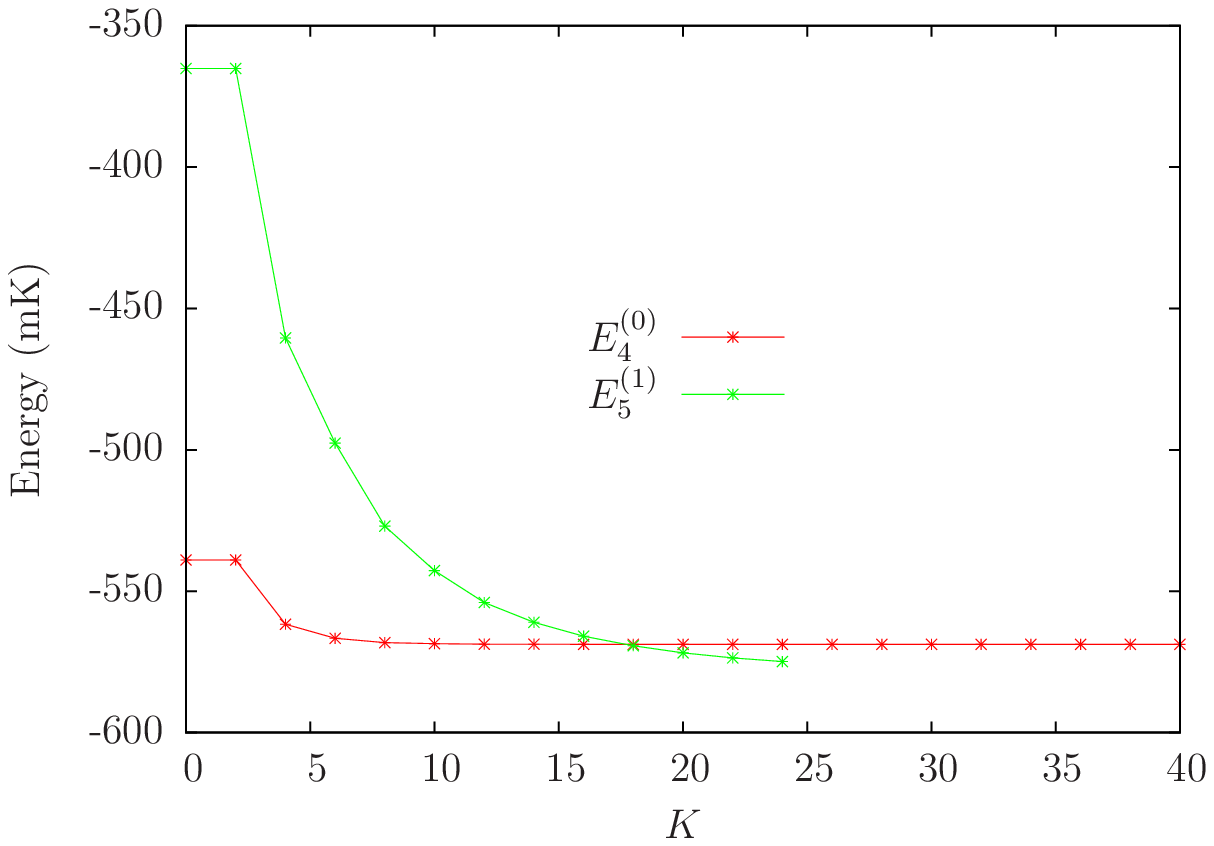}
\caption{The trimer bound state and tetramer first excited state
(left panel) and tetramer bound state and pentamer first excited state
 (right panel) as a functions of $K$.}
\label{fig:A3F2}
\end{figure}

\begin{figure}[h]
  \includegraphics[width=0.5\linewidth]{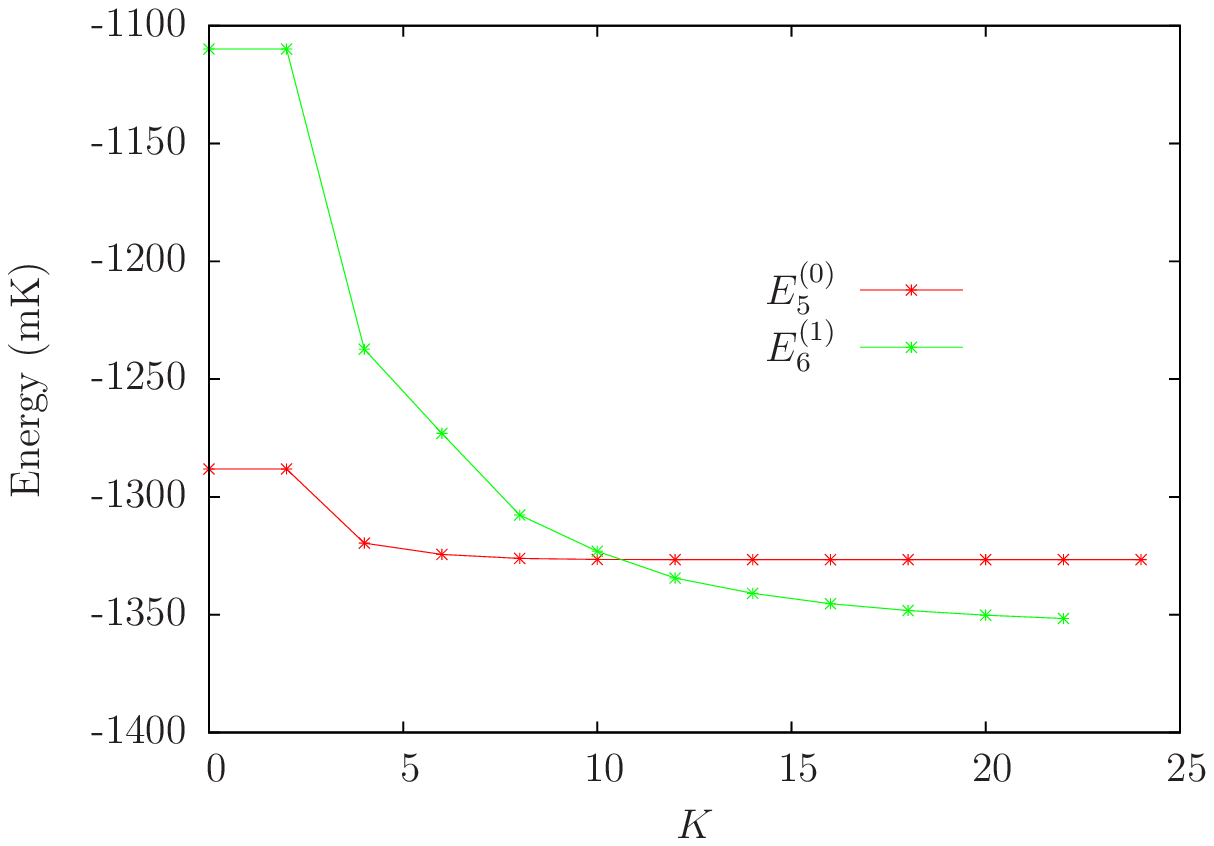}
  \includegraphics[width=0.5\linewidth]{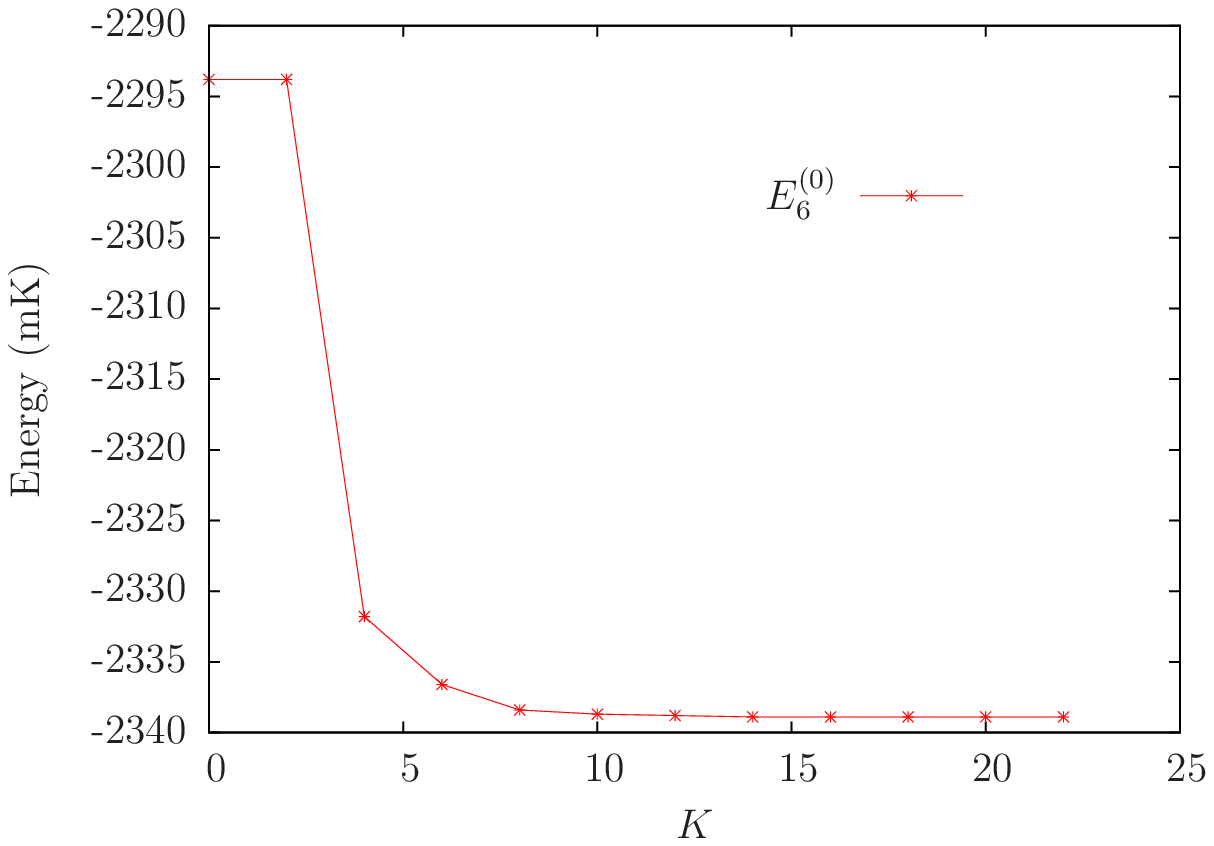}
\caption{The pentamer bound state and hexamer first excited state
(left pannel) and hexamer bound state (right pannel) as a functions of $K$.}
\label{fig:A4F2}
\end{figure}

The results are collected in Figs.~\ref{fig:A3F2},\ref{fig:A4F2} where we
show the convergence in terms of $K$ of the ground state and first excited
state of the bosonic helium clusters. Starting from $A=3$ the bosonic spectrum
is formed by two states, one deep and one shallow close to the threshold
formed by the $A-1$ system with one atom far away. 
The calculations have been performed up to $K=40$ in $A=4$, $K=24$ in $A=5$ and
$K=22$ in $A=6$.  From the figure we can observe that the ground state
binding energy, $E^{(0)}_{A}$, has a very fast convergence in terms of $K$. 
The convergence of the
$E^{(1)}_{A}$ is much slower than for the ground state, however with the
extended based used it has been determined with an accuracy well below $1\%$.
The results confirm
previous analyses in the four body sector that the lower Efimov state in the
$A=3$ system produces two bound states, one deep and one shallow.  Here, we
have extended this observation up to the $A=6$ system. Specifically we have 
obtained the following ground state energies: $E^{(0)}_4=568.8$ K, 
$E^{(0)}_5=1326.6$ K,and $E^{(0)}_6=2338.9$ K, and first excited state
energies: $E^{(1)}_4=129.0$ K, $E^{(1)}_5=574.9$ K,and $E^{(0)}_6=1351.6$ K.
It is interesting to compare the results obtained using the soft-core
representation of the LM2M2 potential with the results of
Refs.~\cite{lewerenz:1997_j.chem.phys.,blume:2000_j.chem.phys.}
obtained using the original LM2M2 interaction. For the
ground state the agreement is around $2\%$ for $A=4,5$ and around $1\%$ for
$A=6$.  The agreement is worst for the excited state, however the results from
Ref.~\cite{blume:2000_j.chem.phys.} are obtained using approximate solutions of the
adiabatic hyperspherical equations. The recent, and very accurate, results 
of Ref.~\cite{hiyama:2012_phys.rev.a} for $A=4$ ($E^{(0)}_4=558.98$ K and
$E^{(1)}_4=127.33$ K) shows a good agreement in particular for the first 
excited state.

Finally it is possible to analyze the ratios $E^{(1)}_{A}/E^{(0)}_{A-1}$.
In the case of Efimov physics these ratios present and universal character.
The He-He potential it is not located exactly at the unitary limit
(infinite value of $a_0$) but it is close to it. 
Using the soft potential models these ratios are:
$E^{(1)}_{4}/E^{(0)}_{3}=1.020$, 
$E^{(1)}_{5}/E^{(0)}_{4} =1.011$ and
$E^{(1)}_{6}/E^{(0)}_{5}=1.018$.  The ratios between
the trimer ground state and the ground states of the $A=4,5,6$ systems are
$E^{(0)}_{4}/E^{(0)}_{3}=4.5$, $E^{(0)}_{5}/E^{(0)}_{3}= 10.5$ and
$E^{(0)}_{5}/E^{(0)}_{3}=18.5$, respectively. These ratios are in good
agreement with those given in
Refs.~\cite{von_stecher:2009_natphys,deltuva:2010_phys.rev.a,von_stecher:2010_j.phys.b:at.mol.opt.phys.}.

The overall agreement of the $A=4,5,6$ systems between LM2M2 and the soft
potential model gives a further indication that these systems are a nice
realization of which it is called Efimov physics.

\section{Conclusions}\label{sec:conclusions}

In this work we have shown results using the HH expansion in the description
of a $A$-body system with $A=3-6$. The basis has not been symmetrized or
antisymmetrized as required by a system of identical particles. However, the
eigenvectors of the Hamiltonian have well defined permutation symmetry. 
The benefit of the direct use of the HH basis is based on a particular simple 
form used to represent the potential energy. 
We have limited the analysis to consider a central potential. In a first
example we have describe a system of nucleons interacting through the Volkov
potential, used several times in the literature.
Though the use of a central potential leads to an unrealistic description of the
light nuclei structure, the study has served to analyze the
characteristic of the method: the capability of the diagonalization
procedure to construct the proper symmetry of the state and the particular structure,
in terms of products of sparse matrices, of the Hamiltonian matrix. The
success of this study makes feasible
the extension of the method to treat interactions depending on spin and isospin
degrees of freedom as the realistic NN potentials. A preliminary analysis
in this direction has been done~\cite{gattobigio:2009_few-bodysyst.}.

In a second example we have
studied the possibility of calculating bound and excited states in a bosonic
system consisting of helium atoms interacting through soft two- and three-body 
forces. The potential model has been adjusted to approximate the description of 
small helium clusters interacting through one of the realistic helium-helium
interactions, the LM2M2 potential. After the
direct diagonalization of the $A$-body system we have observed that clusters
with $A=3,4,5,6$ atoms present a deep bound state and a shallow bound state
just below the energy of the $A-1$ system. 
Since the He-He potential predicts a large two-body scattering length we have
studied the universal ratios $E^{(1)}_{A}/E^{(0)}_{(A-1)}$.
These ratios have been studied in detail in
the $A=4$ case (see Refs.~\cite{hammer:2007_eur.phys.j.a,deltuva:2010_phys.rev.a}).
Estimates have been obtained also for bigger
systems~\cite{von_stecher:2010_j.phys.b:at.mol.opt.phys.}. Our calculations, obtained for one
particular value of the ratio $r_0/a$, are in agreement with those ones. An
analysis of the universal ratios as $a\rightarrow\infty$ is at present under
way.

\end{document}